\documentclass[twocolumn,showpacs,preprintnumbers,prd,superscriptaddress,
nofootinbib]{revtex4}
\usepackage{comment}
\usepackage{graphicx}
\usepackage{amsmath, mathrsfs}
\usepackage{float}
\usepackage{subfigure}
\usepackage[usenames]{color}
\usepackage{amssymb}
\usepackage{bbm}
\usepackage{array}    
\usepackage{csquotes}
\usepackage{color}
\usepackage{psfrag}
\usepackage{graphicx}
\usepackage{caption,subcaption}
\usepackage{marginnote}
\usepackage[utf8]{inputenc}
\usepackage{tikz}
\usetikzlibrary{positioning,decorations.pathmorphing}

\usepackage{epstopdf}
\newcommand{\bea}{\begin{aligned}}
\newcommand{\eea}{\end{aligned}}
\newcommand{\beq}{\begin{equation}}
\newcommand{\eeq}{\end{equation}}

\newcommand{\bse}{\begin{subequations}}
\newcommand{\ese}{\end{subequations}}

\usepackage{hyperref}
\hypersetup{
     colorlinks   = true,
     citecolor    = red,
     linkcolor    = blue,
     urlcolor     = blue,
}

\newcommand{\bmm}{\begin{multline}}
\newcommand{\emm}{\end{multline}}

\begin{document}

\begin{center}    
\end{center}
\title{Effective matter sectors from modified  entropies}
\begin{center}   
\end{center}

\author{Ankit Anand}
\email{anand@iitk.ac.in} 
\affiliation{Department of Physics, Indian Institute of Technology, Kanpur 
208016, India}
\author{Sahil Devdutt}
\email{devduttsahil@gmail.com}  
\affiliation{Department of Physics \& Astronomical Science, Central University 
of Himachal Pradesh, Dharamshala- 176215, India. }
\author{Kimet Jusufi}
\email{kimet.jusufi@unite.edu.mk}
\affiliation{Physics Department, State University of Tetovo, Ilinden Street nn, 
1200, Tetovo, North Macedonia.}

\author{Emmanuel N. Saridakis} \email{msaridak@noa.gr}
\affiliation{National Observatory of Athens, Lofos Nymfon, 11852 Athens, Greece}
\affiliation{Departamento de Matem\'{a}ticas, Universidad Cat\'{o}lica del 
Norte, Avda. Angamos 0610, Casilla 1280 Antofagasta, Chile}
\affiliation{CAS Key Laboratory for Researches in Galaxies and Cosmology, 
School 
of Astronomy and Space Science, University of Science and Technology of China, 
Hefei, Anhui 230026, China}


\begin{abstract}
We present a general formalism linking modified entropy functions directly to a 
modified spacetime metric and, subsequently, to an effective matter sector of 
entropic origin. In particular, within the framework of general relativity, 
starting from the first law of black-hole thermodynamics we establish an 
explicit correspondence between the entropy derivative and the metric function, 
which naturally leads to an emergent stress-energy tensor representing an 
anisotropic effective fluid. This  backreaction effect of horizon entropy may 
resolve possible inconsistencies recently identified in black hole physics 
with 
modified entropies. As specific examples, we apply this 
procedure  to a wide class of modified entropies, such as  
  Barrow, Tsallis-Cirto, Rényi, Kaniadakis, logarithmic, power-law, 
loop-quantum-gravity, and exponential modifications, and we   
derive the associated effective matter sectors, analyzing their physical 
properties and energy conditions. 

\end{abstract}

\maketitle
\section{Introduction}

The connection between gravity, thermodynamics, and quantum theory proves 
to be useful in studying the structure of spacetime. Since the pioneering 
works of Bekenstein and Hawking revealed that black holes behave as 
thermodynamic objects with well-defined temperature and entropy 
\cite{Bekenstein:1973ur, Hawking:1975vcx}, it has become evident that the laws 
of gravity are related to a   thermodynamic structure. This feature 
suggests that spacetime itself may possess microscopic degrees of freedom, and 
that gravitational dynamics could emerge as a macroscopic manifestation of 
their statistical behavior. The area law of black hole entropy and the 
holographic principle  imply that information associated with a volume of space 
can be effectively stored on its boundary \cite{tHooft:1993dmi, 
Susskind:1994vu}. Such ideas have inspired a broad paradigm in which gravity 
arises as an emergent, entropic phenomenon rather than a fundamental force, 
linking the geometry of spacetime to the thermodynamic behavior of its 
underlying microscopic constituents \cite{Padmanabhan:2009vy, 
Verlinde:2010hp}.\\

The modifications to  the standard Bekenstein-Hawking entropy play a central 
role in exploring possible extensions of semiclassical gravity. The 
conventional area law, \( S = \tfrac{A}{4} \), represents the leading-order 
contribution arising from quantum fields near the horizon, but various 
approaches to quantum gravity and non-equilibrium statistical mechanics suggest 
the presence of subleading corrections. Among these, logarithmic, power-law and 
exponential corrections have received significant attention, as they naturally 
emerge in loop quantum gravity, quantum geometry, and generalized thermodynamic 
formalisms \cite{Kaul:2000kf, Das:2001ic, Sen:2012dw, Tsallis:1987eu, 
Renyi:1961, Barrow:2020tzx, aa, Nakarachinda:2022gsb, PhysRevE.66.056125, 
PhysRevE.72.036108, 
Volovik2025TsallisCirto}.
Power-law corrections to the Bekenstein-Hawking entropy naturally arise when 
entanglement entropy is computed for quantum fields in excited states rather 
than in the vacuum configuration \cite{Das:2007mj}, and in such cases the 
leading area term receives subleading power-law suppressed contributions that 
depend on the excitation level of the quantum state.
Additionally, non-extensive entropy 
frameworks-such as those based on Tsallis and Rényi formalisms, 
quantum-gravitational corrections such as in Barrow entropy, relativistic 
corrections such as in Kaniadakis entropy, etc, 
provide a 
generalized description of gravitational systems, leading to   
modifications of  geometry and thermodynamics and thus to interesting 
black-hole and cosmological phenomenology
\cite{Lymperis:2018iuz,Saridakis:2020lrg,Moradpour:2018ivi,Nojiri:2019skr,
Abreu:2017hiy, Iqbal:2019ooy,
Nojiri:2019itp,Maity:2019qbv,Geng:2019shx,Lymperis:2021qty,Mohammadi:2021wde,
Telali:2021jju,Hernandez-Almada:2021rjs,Jusufi:2022mir,Komatsu:2022bik,
Zamora:2022cqz,Drepanou:2021jiv,Luciano:2022ely,Jusufi:2021fek,Nojiri:2022dkr,
Luciano:2022knb,
Jizba:2022bfz,Chanda:2022tpk,
Dheepika:2022sio,Saha:2022oph,
Luciano:2023zrx,
Luciano:2023fyr,Teimoori:2023hpv,Naeem:2023ipg,Jalalzadeh:2023mzw,
Basilakos:2023kvk, Naeem:2023tcu, Coker:2023yxr, Lymperis:2023prf,
Saavedra:2023rfq,Nakarachinda:2023jko,Jizba:2023fkp,
Okcu:2024tnw,Jalalzadeh:2024qej,Adhikary:2024sax,Jawad:2024yrv,Sheykhi:2024fya, 
Jizba:2024klq,Huang:2024xqk, 
Ebrahimi:2024zrk,
Trivedi:2024inb, Yarahmadi:2024oqv,
Petronikolou:2024zcj,Karabat:2024trf,Ens:2024zzs,Tsilioukas:2024seh,
Ualikhanova:2024xxe,Shahhoseini:2025sgl,Lymperis:2025vup,Nojiri:2025gkq,
Luciano:2025elo,Luciano:2025hjn,Jusufi:2024rba,Dabrowski:2024qkp,Jusufi:2023ayv,
Jusufi:2025hte,Basilakos:2025wwu,Kotal:2025bof,
Nojiri:2025fiu,Capozziello:2025axh,Abreu:2025etn,Luciano:2025fqg}.
Motivated by these developments, in this work, we investigate 
the implications of such generalized entropy functions, incorporating 
logarithmic \cite{Kaul:2000kf, Sen:2013ns, Carlip:2000nv} and exponential 
corrections~\cite{Medved:2004tp, Nojiri:2005sr, Pourhassan:2017qrq, 
Chatterjee:2020iuf}, on the structure of spherically symmetric black hole 
spacetimes.

On the other hand, in recent years, gravity has been increasingly viewed as an 
emergent phenomenon 
rather than a fundamental interaction, arising from the
thermodynamic features of spacetime \cite{Jacobson:1995ab,Padmanabhan:2009vy}. 
In this context, the notion of an \emph{entropic force} provides a useful 
conceptual framework. An entropic force is an effective macroscopic force that 
arises due to the statistical tendency of a system with many microscopic 
degrees 
of freedom to increase its entropy \cite{Verlinde:2010hp}. The force equation 
depends only on entropy differences and is independent of the details of the 
microscopic dynamics, with no fundamental field mediating it. Typical examples 
include colloidal interactions and osmotic pressure, both governed by entropy 
gradients rather than direct mechanical forces. A well-known example is the 
elasticity of a polymer molecule: when immersed in a heat bath, the polymer 
tends to adopt a randomly coiled configuration since such states maximize 
entropy. Stretching the polymer reduces the number of accessible microstates, 
and the system responds with a restoring force that drives it back toward the 
equilibrium configuration \cite{Callen:1985}. 

Similarly, in gravitational systems, entropy gradients associated with the 
microscopic degrees of freedom of spacetime can give rise to a macroscopic 
force 
that one can interpret  as gravity. Within this picture, black hole 
thermodynamics 
provides a natural setting to explore the connection between entropy, 
temperature, and geometry \cite{Bekenstein:1973ur,Hawking:1975vcx}. Hence, 
modifications to the entropy-area relation   
can lead to deviations from the standard Schwarzschild geometry and generate 
new classes of regular black hole metrics. 

In the present work, we are interested in studying
the effect of  entropy 
modifications on the black-hole metric and consequently  on the effective 
matter sector. We mention that our analysis differs from the one of 
\cite{Nojiri:2022ljp,Elizalde:2025iku}, since in those works
the authors examined whether generalized non-extensive entropies   are 
consistent with the Hawking temperature and  the Arnowitt-Deser-Misner (ADM)  
mass, essentially testing the thermodynamic validity of such entropies within a 
fixed spacetime geometry, while   our  approach reverses the logic and we 
derive the modified spacetime metric itself from the chosen entropy function, 
establishing a direct entropy-geometry correspondence rather than treating 
geometry as given. Hence,  by translating the entropy deformation 
into a modified effective fluid description, our framework provides an explicit 
physical realization of how entropy corrections manifest as new, effective
matter sectors of entropic origin. 
Recently, a Lagrangian reconstruction of entropy deformations 
has been proposed in~\cite{DAgostino:2024sgm},  where power-law modifications of 
the Bekenstein-Hawking entropy are studied from a complementary perspective. We 
mention here that 
the Lagrangian formulation follows a different route, since the entropy itself 
is elevated to a dynamical quantity derived from an underlying variational 
principle, allowing one to construct a consistent action whose extremization 
reproduces the desired non-extensive entropy structure.

The plan of the work is the following. In Section \ref{Effectivemattersector}
 we present the connection of modified entropy relations to a modified metric 
function and then to an effective matter sector of entropic origin. Then, in 
Section  \ref{Applications} we     proceed to specific applications to the 
various   modified  entropy forms that exist in the literature, such as  
Barrow, Tsallis-Cirto, Rényi, Kaniadakis, logarithmic, power-law, 
loop-quantum-gravity, and exponential modifications, 
deriving the associated effective matter sectors, and analyzing their physical 
properties and energy conditions. Finally, Section \ref{Conclusions} is devoted 
to the conclusions.

 \section{Effective matter sector from modified  entropy}
 \label{Effectivemattersector}
 
 In this section we first show how a modified horizon entropy can lead to a 
modified spacetime geometry, and then we show how this modified   
geometry can be interpreted to arise from an effective matter sector of 
entropic origin.

\subsection{Modified spacetime geometry from modified horizon entropy}

 In classical general relativity, 
the usual procedure is to specify an energy-momentum tensor, solve for the 
metric components, and subsequently compute the corresponding horizon entropy. 
On the other hand, ideas such as emergent gravity and the holographic principle 
indicate that gravity may arise from entropic or informational degrees of 
freedom,  implying the role of entropy as a fundamental quantity. Guided by 
this 
viewpoint, we propose a simple approach in which the spacetime geometry is 
obtained directly from the horizon entropy. 

Let us begin by assuming a static, 
spherically symmetric line element
\begin{equation}\label{Metric_ansatz}
ds^2 = -f(r)\,dt^2 + \frac{dr^2}{f(r)} + r^2\,d\Omega^2 \ .
\end{equation}
The event horizon \(r_+\) satisfies \(f(r_+)=0\), and the Hawking temperature 
can be computed using $T = f'(r_+)/4\pi$. 
On the other hand, the first law of black-hole thermodynamics is written as
\begin{equation}\label{First_Law}
dM = T\,dS.
\end{equation}
Based on this expression, one can obtain the black-hole horizon entropy as
\begin{equation}
S=\int \frac{1}{T}\frac{\partial M}{\partial r_+}dr_+ \ .
\end{equation}
Using this relation, it is well known that the entropy of a Schwarzschild black 
hole is proportional to the surface area of its horizon, as given by the 
Bekenstein-Hawking entropy formula $S=A/4=\pi r_+^2$, where $A=4 \pi r_+^2$. 
Conversely, one may follow the inverse procedure: starting from the 
Bekenstein-Hawking entropy, it is straightforward to verify that the 
corresponding metric function reproduces the Schwarzschild solution  
(see Appendix  \ref{Appn:Horizon Thermodynamics and Constraint on 
$f(r)$} for the details) and 
for that \(G_{\mu\nu}=0\). 
Thus,  a natural question arises, namely what 
happens in the general case of non--Bekenstein--Hawking entropies, and if one 
can 
deduce the metric function directly from the modified horizon entropy.

We   consider a general functional dependence of the black-hole 
entropy on the horizon radius, i.e.
\begin{equation}\label{General_Entropy}
    S = S(r_+) \, .
\end{equation}
To construct a metric consistent with this entropy we take the following 
ansatz 
for the metric function
\begin{equation}\label{Metric_ansatz}
    f(r) = 1 - M\,g(r) \, ,
\end{equation}
where $M$ denotes the Arnowitt-Deser-Misner (ADM)  mass of the black hole and 
$g(r)$ is an arbitrary 
function of the radial coordinate. 
The reason for choosing the form \eqref{Metric_ansatz} is that in the limit 
where the entropy reduces to the Bekenstein-Hawking area law, one must recover 
the Schwarzschild metric.  
Therefore, the parametrization \eqref{Metric_ansatz} is the minimal choice that 
ensures a smooth reduction to the Schwarzschild solution in the appropriate 
limit. 

The location of the event horizon is determined from $f(r_+)=0$, which 
gives
\begin{equation}
    M = \frac{1}{g(r_+)} \, .
\end{equation}
The Hawking temperature follows from the surface gravity and is given by
\begin{equation}\label{Temperature_expression}
    T = \frac{f'(r_+)}{4\pi}
    = -\,\frac{g'(r_+)}{4\pi\, g(r_+)} \,,
\end{equation}
where the prime denotes differentiation with respect to $r$ and the expression 
is evaluated at the horizon. Now, using the first law of 
thermodynamics  \eqref{First_Law} and substituting 
relations \eqref{General_Entropy}-\eqref{Temperature_expression},
we obtain
\begin{equation}\label{General_g(r)}
    g(r_+) = \frac{4\pi}{S'(r_+)} \, ,
\end{equation}
where we have considered the non-trivial case where  $g'(r_+)\neq0$ (ensuring 
a simple 
first-order zero of $f(r)$ and hence
a regular, non-extremal event horizon).

Relation \eqref{General_g(r)}, which is the basic point of this work,  encodes 
the backreaction effect of the entropy into the spacetime geometry. As one can 
see,  it has been is derived 
locally at the event horizon. However, we can still use it in order to 
determine the full spacetime geometry, provided that 
we make the following assumptions:\\

 (i) Extending the horizon relation, namely  promoting the horizon relation 
$g(r_+)=4\pi/S'(r_+)$ to a global functional
dependence, $ g(r) = 4\pi/S'(r)$,
which uniquely reconstructs the metric function from the chosen entropy
functional form. This prescription ensures that the thermodynamic input
$S=S(r_+)$ is encoded directly in the geometry, and thus given $S(r)$, the 
metric function 
is fixed everywhere.\\

(ii) Identifying the ADM mass with the parameter $M$, which requires that the 
metric 
asymptotically
reduces to the Schwarzschild form when $S(r)=\pi r^2$. Under the above 
extension,
$S' = 2\pi r$ implies $g(r)=2/r$ and hence $f(r)=1-2M/r$,
confirming that the Schwarzschild solution is recovered when the entropy
reduces to the Bekenstein-Hawking area law.  We emphasize that 
the present construction is limited to static, spherically symmetric spacetimes, 
and thus the proposed entropy-geometry correspondence should be viewed as a 
controlled realization within this symmetry class. Generalizations to rotating 
or dynamical geometries (such as Kerr or time-dependent spacetimes) lie beyond 
the scope of the present work and will be explored in future investigations. 

In summary, under the above assumptions, we can write  
\begin{equation}
    f(r) =  1 - \frac{4\pi M}{S'(r)} 
    \label{Generalfr},
\end{equation}
which shows that once the functional form of the 
entropy   is specified, one can obtain effectively the corresponding metric 
function, and thereby the 
spacetime geometry.   In the special 
case 
where $S(r) = \pi r^2$ (Bekenstein-Hawking entropy), we recover the 
Schwarzschild solution, as expected. We would like to clarify the interpretation 
of the function $S(r)$ in our construction.  
It is important to distinguish between Bekenstein-Hawking (thermodynamic) 
entropy, which is associated with a black hole horizon, and \emph{entanglement 
entropy} of quantum fields across a spherical boundary.  
For radii $r > r_+$, our $S(r)$ should be understood as the \emph{entanglement 
entropy} of vacuum fields across the spherical surface at radius $r$, rather 
than as a thermodynamic entropy of matter contained in the sphere.  
Entanglement entropy counts quantum correlations between degrees of freedom 
inside and outside the surface and is known to satisfy an \emph{area law} 
scaling $S_{\rm ent}(r) \sim r^2.$ The Bekenstein bound, $S \le 2 \pi E r,$
applies only to physical systems where the entropy is realized by energy $E$ (or 
mass $M$) contained inside a radius $r$.  
Since entanglement entropy does not correspond directly to the energy content, 
it is not constrained by this bound.  
Thus, the quadratic growth of $S(r)$ with $r$ does \emph{not} violate any 
fundamental physical principle.
In our metric construction, $S(r)$ encodes the cumulative entanglement entropy 
of vacuum fields. Its role is not thermodynamic, and therefore the extension of 
$S(r)$ to $r > r_+$ is physically meaningful and consistent with semiclassical 
expectations. 
Our $S(r)$ in the region $r_+>r$ is the entanglement entropy and does not scale 
with energy, and the Bekenstein bound cannot be apply to such regions. This also 
suggest that one can modify the Bekenstein bound in order to include the total 
$S(r)=S_{BH}(r_+)+S_{\rm ent}(r_+,r)$ and Bekenstein bound applies only the 
first term, while the second term is not constrained by the Bekenstein bound; it 
is constrained by the area (holographic/covariant) bound that can be written for 
the arbitrary surface 
\begin{equation}
S_{BH}(r_+)+S_{\rm ent}(r_+,r) \leq \frac{A(r)}{4G}
\end{equation}
Although we solved $f(r)$ explicitly at the horizon, the solution can be 
extended to arbitrary $r$ by invoking the continuity and smoothness of the 
metric function, together with the boundary conditions imposed at the horizon. 
In this way, the resulting metric satisfies the Einstein field equations for all 
$r$ outside the horizon. The metric construction in our paper is therefore 
physically consistent.

In addition, our approach is consistent with uniqueness, namely the global 
extension does not violate any physical constraint: Birkhoff's theorem ensures 
that any spherically symmetric vacuum solution outside the horizon is unique.  
Therefore, the choice of $g(r)$ is not arbitrary, but is in fact constrained by 
the requirement that the solution remain the unique static, spherically 
symmetric vacuum geometry compatible with the given boundary conditions at 
infinity.  
In other words, our ``entropy--geometry correspondence'' is compatible with the 
uniqueness of the Schwarzschild solution, and the functional form of $g(r)$ is 
consistent with Birkhoff's theorem. In the case of Bekenstein-Hawking entropy we 
do get vacuum solution, however deviations from Bekenstein-Hawking entropy can 
be interpreted as non-vacuum solution, or contributions from contributions 
effective matter sectors. 

 We proceed by  showing how in our setup the 
gravitational 
force emerges  from the horizon entropy. From the relation 
\begin{equation}
    f(r) \equiv 1+2 \phi_G\, ,
\end{equation}
where
$
    \phi_G \equiv - \frac{2\pi M}{S'(r)}
$
 is the gravitational potential,
we can obtain the gravity force acting on a test particle with mass $m$ near 
$M$ 
using
\begin{equation}
    \vec{F}_G=-m \nabla \phi_G =  2\pi m M \nabla  \left(\frac{1}{S'(r)}\right) 
\hat{r},
\end{equation}
  obtaining the universal law of gravity as
\begin{equation}
    \vec{F}_G=  -2\pi m M \frac{S''(r)}{S'(r)^2}\hat{r}.
\end{equation}
In other words, the attractive force of gravity can be viewed as an emergent 
effect arising from the change of horizon entropy. This is in line to the 
entropic force scenario proposed by Verlinde~\cite{Verlinde:2010hp}.

Let us now assume a general entropy relation, which can be expressed as a 
correction to Bekenstein--Hawking expression, namely
\begin{equation}
S = S_{BH}+\mathcal{S}(A).
\end{equation}
Then using 
\begin{equation}
\frac{\partial S}{\partial r}=\frac{\partial S}{\partial A}\frac{\partial 
A}{\partial r}=\left(\frac{1}{4}+\frac{\partial \mathcal{S}}{\partial A}  
\right)8 \pi r
\end{equation}
and 
\begin{equation}
\frac{\partial^2 S}{\partial r^2}=2\pi \left(1+4  \frac{\partial 
\mathcal{S}}{\partial A}\right)+8 \pi r  \frac{\partial }{\partial 
r}\left(\frac{\partial \mathcal{S}}{\partial A}\right),
\end{equation}
for the universal law of gravity we finally acquire
\begin{equation}\label{New}
    \vec{F}_G=  - \frac{m M}{r^2} \left[\frac{ \left(1+4  \frac{\partial 
\mathcal{S}}{\partial A}\right)+ 4 r\frac{\partial }{\partial 
r}\left(\frac{\partial \mathcal{S}}{\partial 
A}\right)}{\left(\frac{1}{4}+\frac{\partial \mathcal{S}}{\partial A}  \right)^2 
16  }\right]  \hat{r}.
\end{equation}
Note that  if the entropy is just the Bekenstein-Hawking entropy,  we reproduce
  the Newton's law force, i.e.
$
    \vec{F}_G=  - \frac{m M}{r^2}\hat{r}$. Hence,
from Eq. \eqref{New} we see that a deviation from the Bekenstein-Hawking entropy 
can be interpreted as a 
modified gravitational law of gravity.

\subsection{Effective matter sector of entropic origin}

In this subsection we show how a modified  entropy relation can be interpreted 
to 
lead to an effective matter sector.
Let us now use the metric (\ref{Generalfr}) to calculate the components of the  
Einstein tensor $G^{\mu}_{\ \nu}$. We find 
\begin{equation}
\label{einsteintt}
    G_t^t =G_r^r = \frac{4 \pi  M \left(r S''(r)-S'(r)\right)}{r^2 S'(r)^2},
\end{equation}
and 
\begin{equation}
\label{einsteinthth}
    G_\theta^\theta = G_\phi^\phi = \frac{2 \pi  M \left\{S'(r) [r 
S^{'''}(r)+2 S''(r)]-2 r S''(r)^2\right\}}{r S'(r)^3}\ ,
\end{equation}
which  are 
non-zero in general for an entropy relation different from Bekenstein-Hawking 
one. Hence, from 
the  field equations of general relativity
\begin{eqnarray}
G^{\mu}_{\ \nu} = 8\pi\, T^{\mu}_{\ \nu},
\end{eqnarray}
we conclude that we obtain a non-zero,  effective   stress-energy tensor  
of entropic origin. In particular, we can define
$T^{\mu}_{\nu}=\left(-\rho,  p_{r},  p_{t},  p_{t}  \right)$, with
\begin{align}
\label{rhoeq}
    \rho(r) &= -\frac{M\left[r S''(r) - S'(r)\right]}{2\, r^{2} 
S'(r)^{2}},\\[4pt]
    p_{r}(r) &= -\rho(r),
    \label{preq}
    \\[4pt]
    p_{t}(r) &= \frac{M\left\{S'(r)[r S^{'''}(r) + 2 S''(r)] - 2 r 
S''(r)^{2}\right\}}{4\, r\, S'(r)^{3}}.
    \label{ptreq}
\end{align}
Note that the relation \(p_{r} = -\rho\) reflects a vacuum-like or 
dark-energy-type equation of state in the radial direction. However, since 
\(p_{t} \neq p_{r}\), this effective matter sector of entropic origin is 
anisotropic. Thus, the deviation of the entropy function 
\(S(r)\) from the standard area law acts as a geometric source generating an 
anisotropic stress-energy tensor without introducing any explicit matter 
fields. In this sense, the modified entropy behaves as an effective, emergent 
gravitational matter content associated with horizon microstructure.  This is 
the main result of the present work. In the following we apply it for the known 
modified entropy relations of the literature.

It is important to emphasize that the inverse thermodynamic construction adopted 
here does not aim to generate arbitrary vacuum geometries. Rather, once a 
modified entropy functional is specified, the resulting metric is determined by 
thermodynamic consistency and generally corresponds to a non-vacuum 
configuration. The associated stress-energy tensor obtained from the Einstein 
equations should therefore be interpreted as an effective matter sector encoding 
the corrections implied by the chosen entropy. In this sense, the framework 
establishes a mapping between entropy modifications and emergent matter content.

Next, let us show the conservation of the energy-momentum tensor for the general 
case in our setup. Let us start from the relation
\begin{eqnarray}
    \nabla_\mu  T^{\mu \nu}=0,
\end{eqnarray}
from this relation one can obtain 
\begin{eqnarray}
    \frac{d p_r}{dr}=-\frac{1}{2 
g_{tt}}\frac{dg_{tt}}{dr}(\rho+p_r)+\frac{2}{r}(p_t-p_r).
\end{eqnarray}
From the last equation and using the condition $\rho=-p_r$, it follows that
\begin{eqnarray}\label{p_trhodrho}
   p_t=-\rho-\frac{r}{2} \frac{d \rho}{dr}.
\end{eqnarray}
Using Eq.~\eqref{rhoeq} and Eq.~\eqref{p_trhodrho} we indeed get 
Eq.~\eqref{ptreq} as follows
\begin{align}\notag
   p_t &=\frac{M\left[r S''(r) - S'(r)\right]}{2\, r^{2} 
S'(r)^{2}}-\frac{r}{2} \frac{d}{dr}\left[-\frac{M\left[r S''(r) - 
S'(r)\right]}{2\, r^{2} 
S'(r)^{2}}\right] \\
&=\frac{M\left\{S'(r)[r S^{'''}(r) + 2 S''(r)] - 2 r 
S''(r)^{2}\right\}}{4\, r\, S'(r)^{3}}.
\end{align}
This  result shows that the conservation of energy-momentum tensor holds in our 
setup. 

 We mention here that the effective stress-energy tensor derived from the 
reconstructed geometry can be interpreted as an anisotropic fluid, characterized 
by distinct radial and tangential pressures. Such effective matter sectors 
commonly arise in semiclassical gravity and quantum-corrected spacetimes. 
Violations of classical energy conditions in certain parameter regimes are 
therefore not necessarily pathological, but rather indicative of underlying 
quantum or statistical corrections to classical gravitational dynamics.
  
\section{Application to specific modified entropy}
 \label{Applications}

 In the previous sections we showed how a modified entropy expression leads to 
an effective matter sector. Hence, we can now proceed to application to the 
various specific entropy forms that exist in the literature.

\subsection{Barrow Entropy}

Barrow argued that quantum gravitational corrections may change the classical 
smoothness of the event horizon, giving rise to a horizon geometry with fractal 
characteristics. Such a modification implies that the standard area law for 
black hole entropy does not hold exactly. To quantify the degree of this 
geometric irregularity, a parameter $\Delta$ is introduced, representing the 
extent to which the horizon departs from a smooth two-dimensional surface. With 
this modification, the entropy associated with a black hole is expressed as   
\cite{Barrow:2020tzx} 
\begin{equation}\label{SB}
S_B = \left( S_{BH} \right)^{1+\frac{\Delta}{2}} \ ,
\end{equation}
where $0 \leq \Delta \leq 1$. The case $\Delta = 0$ corresponds to an 
undeformed 
horizon and reproduces the standard Bekenstein-Hawking entropy, while nonzero 
values of $\Delta$ encode the influence of quantum-gravity-induced fractal 
structure. Conceptually, the presence of a nonzero $\Delta$ indicates that the 
microstructure of spacetime at the horizon deviates from classical smoothness, 
potentially reflecting underlying quantum gravitational degrees of freedom. 
Thus, Barrow entropy provides an effective way to model such corrections 
without 
specifying the detailed microscopic theory.

Using  \eqref{Generalfr}, the Barrow corrected metric function is 
\begin{equation}
    f_{B}(r) = 1-\frac{4 M}{\sqrt{\pi ^\Delta}(\Delta +2)r^{\Delta +1}} \ ,
\end{equation}
the corresponding Einstein tensor components 
(\ref{einsteintt})-(\ref{einsteinthth}) are 
\begin{eqnarray}
    G_t^t &=&G_r^r = 
 \frac{4 \pi ^{-\frac{\Delta }{2}} \Delta  M r^{-\Delta -3}}{\Delta +2},\\  
G_\theta^\theta &=& G_\phi^\phi = -\frac{2 \pi ^{-\frac{\Delta }{2}} \Delta  
(\Delta +1) M r^{-\Delta -3}}{\Delta +2} \ ,
\end{eqnarray}
and thus  the effective anisotropic matter sector
(\ref{rhoeq})-(\ref{ptreq}) becomes 
\begin{equation}
T^{\mu}{}_{\nu}
 = 
\rho\,\mathrm{diag}\!\left(-1,\,-1,\,\tfrac{\Delta+1}{2},\,\tfrac{\Delta+1}{2}
\right) \ ,
\end{equation}
where
\begin{equation}
\rho = -\frac{\pi^{-\frac{\Delta}{2}}\Delta M\,r^{-\Delta-3}}{2\pi(\Delta+2)}
\ .
\end{equation}
 In the limit $\Delta \to 0$, the effective stress tensor vanishes and the 
spacetime reduces to the Schwarzschild vacuum.
Finally, note the gravitational force \eqref{New} for this case reads as
\begin{equation}
    \vec{F}_G=  -2\pi M \frac{S_B''(r)}{S_B'(r)^2}\hat{r}= -\frac{2M (\Delta 
+1) 
}{\sqrt{\pi ^{\Delta}}\;(\Delta+2)r^{\Delta+2}} \hat{r}.
\end{equation}

Let us briefly examine the energy conditions. 
For $M>0$ and $\Delta>0$, 
the 
effective energy density is negative ($\rho<0$). Since \(\rho+p_r = 0\) along 
the 
radial null direction, the null energy condition (NEC) is saturated. Along the 
tangential direction we have
\(\rho+p_t = \tfrac{\Delta+3}{2}\rho < 0\), hence the NEC is violated. Since 
$\rho<0$, the weak energy condition (WEC) is violated. Additionally, since
$\rho+p_r+2p_t=(\Delta+1)\rho<0$,  the strong  energy condition (SEC) is 
violated. Lastly, since $\rho\ge0$ and $|p_i|\le\rho$,  the dominant  energy 
condition  (DEC) is violated, too. Thus, only the radial component marginally 
satisfies the NEC, while all other standard energy conditions are violated for 
$\Delta>0$.

In summary, as we observe the effective matter sector behaves as an 
anisotropic fluid with negative 
energy density equal to the radial pressure, while the tangential pressure 
differs by a factor $(\Delta+1)/2$. Such stress-energy forms cannot arise from 
ordinary classical matter, and they  reflect the  quantum-gravitational or 
fractal  corrections encoded by the Barrow entropy modification. The violation 
of the standard energy conditions is therefore not pathological but signals the 
presence of an effective, non-classical source required to support the modified 
horizon geometry.

\subsection{Tsallis entropy}

The Tsallis-Cirto entropy represents a non-additive extension of the standard 
Bekenstein-Hawking entropy, inspired by the formalism of non-extensive 
statistical mechanics. In this framework, the entropy-area relation is modified 
to accommodate possible correlations or long-range interactions among the 
microscopic degrees of freedom associated with the horizon. This generalized 
entropy has been applied in gravitational and cosmological contexts, 
particularly in approaches where gravitational dynamics emerge from underlying 
thermodynamic principles. Within such scenarios, the Tsallis-Cirto entropy 
leads 
to modified cosmological evolution equations, offering an alternative route to 
explaining late-time cosmic acceleration and the effective behavior attributed 
to dark energy. For a black-hole horizon, the 
Tsallis-Cirto entropy is expressed as \cite{Tsallis:1987eu}
\begin{equation}
S_{TC} = \left(S_{BH}\right)^{\delta} \ ,
\end{equation}
where $\delta$ denotes the non-extensive deformation parameter. The classical 
Bekenstein-Hawking entropy is recovered in the limit $\delta \to 1$, signifying 
the absence of non-extensive effects and a return to the standard area law. The 
parameter $\delta$ quantifies the degree to which horizon degrees of freedom 
are 
correlated or interacting at long ranges. 

The corrected metric function   \eqref{Generalfr} becomes
\begin{eqnarray}
    f_{TC}(r) = 1-\frac{2M  \pi ^{1-\delta }}{\delta \, r^{2\delta-1 }} \ ,
\end{eqnarray}
while the Einstein tensor components (\ref{einsteintt})-(\ref{einsteinthth}) 
 read
\begin{eqnarray}
    G_t^t &=& G_r^r = \frac{4M \pi ^{1-\delta } (\delta -1) }{\delta \; 
r^{2\delta +1}} \\  G_\theta^\theta &=& G_\phi^\phi = -\frac{2 M\pi ^{1-\delta 
} 
(\delta -1) (2 \delta -1) }{\delta \; r^{2\delta +1}} \ .
\end{eqnarray}
Moreover, 
the gravitational force  \eqref{New}  for this case reads
\begin{equation}
    \vec{F}_G=  -2\pi M \frac{S_{TC}''(r)}{S_{TC}'(r)^2}\hat{r}= 
-\frac{M\pi^{1-\delta}(2 \delta -1)}{r^{2\delta} \;\delta } \hat{r} \ .
\end{equation}
Lastly, the effective stress-energy tensor (\ref{rhoeq})-(\ref{ptreq}) becomes
\begin{equation}
T^{\mu}{}_{\nu} 
=\rho\,\mathrm{diag}\!\left(-1,\,-1,\,\tfrac{2\delta-1}{2},\,\tfrac{2\delta-1}{2
}\right),
\end{equation}
where
\begin{equation}
T^{\mu}{}_{\nu} 
=\rho\,\mathrm{diag}\!\left(-1,\,-1,\,\tfrac{2\delta-1}{2},\,\tfrac{2\delta-1}{2
}\right),
\end{equation}
with 
\begin{equation}
\rho=-\frac{M\,\pi^{-\delta}(\delta-1)}{2\,\delta\,r^{2\delta+1}} \ .
\end{equation}

As we observe the Tsallis exponent $\delta$ quantifies also the behavior of the 
effective matter sector. For $\delta=1$ we reproduce the vacuum 
limit ($T^{\mu}{}_{\nu}=0$), while   $\delta\neq1$ describes an 
anisotropic effective matter source. The sign of $\rho$ depends on 
$(\delta-1)$, 
i.e.  for $\delta>1$, $\rho<0$ the effective source has negative 
energy density, while for $0<\delta<1$, $\rho>0$ we obtain a positive 
anisotropic matter distribution.

Concerning the energy conditions, we can see that radial NEC is saturated, but 
the tangential NEC \(\rho+p_t=\tfrac{2\delta+1}{2}\rho\) is satisfied for 
$\rho>0$ and violated for 
$\rho<0$. Moreover, the  WEC holds if $\rho>0$ (i.e.\ $\delta<1$), otherwise 
it is violated. Concerning SEC we find that 
\(\rho+p_r+2p_t=(2\delta-1)\rho\), and thus it is satisfied for $\rho>0$ and 
$\delta>\tfrac{1}{2}$, while it is violated otherwise. DEC holds only if 
$\rho>0$ and $|p_i|\le\rho$, which restricts $\tfrac{1}{2}\le\delta\le1$. 
Hence, 
the stress tensor satisfies all standard energy conditions for $0<\delta<1$ but 
violates them when $\delta>1$, where $\rho$ becomes negative.

In summary, the parameter $\delta$ controls the power-law behavior of the 
effective energy 
density, $\rho\propto r^{-(2\delta+1)}$, producing an anisotropic fluid with 
radial tension $p_r=-\rho$ and tangential pressure proportional to 
$(2\delta-1)\rho/2$. For $\delta<1$ the matter distribution is physically 
reasonable and satisfies the energy conditions, whereas for $\delta>1$ the 
energy density becomes negative, indicating an exotic effective source required 
to sustain a regularized or non-classical geometry. In the   limit 
$\delta\to1$, all stress components vanish and the spacetime smoothly reduces 
to the Schwarzschild vacuum.

\subsection{Renyi Entropy}

Rényi entropy offers a generalized measure of entropy that extends beyond the 
additive structure of the Bekenstein-Hawking formulation. The key feature of 
this framework is the parameter $\lambda$, which controls the degree to which 
the entropy departs from extensivity. Such a modification is useful in 
black-hole 
thermodynamics since the microscopic degrees of freedom associated with 
the horizon may interact in ways that are not accounted for by standard 
Boltzmann-Gibbs statistics.

Interpreting $\lambda$ as an additional thermodynamic parameter enlarges the 
black-hole phase space and enables a modified form of the first law and Smarr 
relation. This approach also leads to changes in the thermodynamic behavior and 
stability properties of black holes when compared to the usual area law 
\cite{aa,Nakarachinda:2022gsb}. The Rényi entropy is given by \cite{Renyi:1961}
\begin{equation}
S_R = \frac{\log\!\left(1 + \lambda \, S_{BH}\right)}{\lambda} \ ,
\end{equation}
and reduces to the usual Bekenstein-Hawking entropy in the limit $\lambda \to 
0$, indicating that the classical area law is recovered when no non-extensive 
effects are present.

The metric function is 
\begin{eqnarray}
    f_R(r) = 1-\frac{2M \left(1+ \pi  \lambda  r^2\right)}{r} \ ,
\end{eqnarray}
the Einstein tensor components are 
\begin{eqnarray}
    G_t^t &=&G_r^r = -\frac{4 \pi  \lambda  M}{r},  \\
    G_\theta^\theta &=& G_\phi^\phi = -\frac{2 \pi  \lambda  M}{r},
\end{eqnarray}
and the gravitational force becomes
\begin{equation}
    \vec{F}_G=  -2\pi M \frac{S_R''(r)}{S_R'(r)^2}\hat{r}= M \left(\pi  \lambda 
-\frac{1}{r^2}\right)\hat{r} \ .
\end{equation}
Additionally, the corresponding effective stress-energy tensor 
(\ref{rhoeq})-(\ref{ptreq}) is
\begin{equation}
T^{\mu}{}_{\nu} = 
\rho\,\mathrm{diag}\!\left(-1,\,1,\,\tfrac{1}{2},\,\tfrac{1}{2}\right), \quad 
\text{where}\quad \rho = \frac{\lambda M}{2r} \ .
\end{equation}
In the limit $\lambda\to0$, one recovers $T^{\mu}{}_{\nu} \to 0$, i.e. the 
Schwarzschild vacuum. Finally, note that  for $M>0$ and $\lambda>0$, the 
effective energy density is positive ($\rho>0$), and all standard energy 
conditions are therefore satisfied for $\lambda>0$.

In summary, the effective matter sector behaves as an anisotropic fluid with 
positive energy density $\rho\propto1/r$, a radial pressure equal to the energy 
density ($p_r=\rho$), and a smaller tangential pressure ($p_t=\rho/2$). This 
represents a non-vacuum configuration sustained by an extended, inhomogeneous 
distribution rather than a delta-function source. The parameter $\lambda$ 
controls the strength of the deviation from the Schwarzschild vacuum, thus
$\lambda\to0$ restores vacuum geometry, whereas finite $\lambda$ introduces a 
mild, physically reasonable anisotropy consistent with all energy conditions.

\subsection{Kandiakis Entropy}

Kaniadakis proposed a generalized statistical framework that departs from the 
traditional Boltzmann-Gibbs formulation by introducing a deformation parameter 
$\kappa$. This approach, often referred to as Kaniadakis statistics, is 
constructed 
to be compatible with relativistic dynamics while maintaining the foundational 
consistency of standard statistical mechanics \cite{PhysRevE.66.056125, 
PhysRevE.72.036108}. Within this framework, the usual Maxwell-Boltzmann 
distribution arises as a special limiting case, whereas nonzero values of 
$\kappa$ 
encode deviations associated with generalized thermodynamic behavior.

When applied to gravitational systems, particularly black holes, this modified 
entropy provides a natural way to incorporate  corrections to horizon 
thermodynamics. The corresponding Kaniadakis entropy for 
black holes takes the form
\begin{equation}
S_K = \frac{\sinh\!\left(\kappa S_{BH}\right)}{\kappa}  \ ,
\end{equation}
where $S_{BH}$ is the standard Bekenstein-Hawking entropy. In the limit $\kappa 
\rightarrow 0$, the expression reduces smoothly to $S_{BH}$, demonstrating that 
the Kaniadakis framework contains the conventional entropy law as a special 
case.

The parameter $\kappa$ can be interpreted as a measure of deviations from the 
standard thermodynamic behavior encoded by the standard horizon geometry. A 
nonzero value of $\kappa$ reflects the presence of additional microscopic 
correlations or fluctuations that are not captured by the ordinary 
Bekenstein-Hawking description. In this sense, the Kaniadakis entropy provides 
an effective macroscopic signature of   statistical features 
near the event horizon, while still preserving continuity with conventional 
black-hole thermodynamics when $\kappa \to 0$.

For this modified entropy the corrected metric function becomes
\begin{equation}
    f_\kappa(r) = 1-\frac{2 M \;\text{sech}\left(\pi \, \kappa \, 
r^2\right)}{r},
\end{equation}
the Einstein tensor components are 
\begin{eqnarray}
    G_t^t &=&G_r^r = \frac{4 \pi  \kappa  M \tanh \left(\pi  \kappa  r^2\right) 
\text{sech}\left(\pi  \kappa  r^2\right)}{r}\\  
G_\theta^\theta &=& \frac{\pi  \kappa  M  \left(\sinh \left(2 \pi  \kappa  
r^2\right)-2 \pi  \kappa  r^2 \left(\cosh \left(2 \pi  \kappa  
r^2\right)-3\right)\right)}{r\;\cosh^3\left(\pi  \kappa  r^2\right)} \ , 
\nonumber
\end{eqnarray}
and the gravitational force  reads
\begin{equation}
    \vec{F}_G =  -\frac{M \left(2 \pi  \kappa  r^2 \tanh \left(\pi  \kappa  
r^2\right)+1\right) \text{sech}\left(\pi  \kappa  r^2\right)}{r^2}\, \hat{r} \ .
\end{equation}
Furthermore, the  corresponding effective stress-energy tensor is
\begin{eqnarray}
\rho &=& -\frac{\kappa M}{2r}\,
\tanh(\pi\kappa r^2)\,\text{sech}(\pi\kappa r^2), \\[3pt]
p_r &=& -\rho, \\[3pt]
p_t &=& \frac{\kappa M}{8r}\,
\frac{\sinh(2\pi\kappa r^2)-2\pi\kappa r^2\bigl(\cosh(2\pi\kappa r^2)-3\bigr)}
{\cosh^3(\pi\kappa r^2)} .\ \ \ \ 
\end{eqnarray}
Hence, the effective matter distribution is anisotropic, with $p_r=-\rho$, and 
the
$t$-direction pressure $p_t$ is determined by the hyperbolic functions of 
$\pi\kappa r^2$.  The parameter $\kappa$ controls the strength of the deviation 
from the vacuum geometry, and for    $\kappa \to 0$ we find 
$T^{\mu}{}_{\nu}\to0$ and the Schwarzschild limit is recovered.

Concerning the energy conditions, we can see that at small $r$  we have 
\(\rho\simeq -\tfrac{\pi\kappa^2 M r}{2}<0\). Thus, near the origin  $\rho<0$ 
and the energy conditions are violated, while for intermediate $r$ the sign of 
$\rho$ may change depending on $\kappa$. The radial NEC is saturated, but the 
tangential NEC can be negative near the core, indicating local NEC violation 
that softens the central singularity. Moreover, WEC and SEC are violated 
wherever $\rho<0$, and they are satisfied only in outer regions where 
$\rho>0$. DEC is generally violated near the 
center due to the negative $\rho$. Hence, the effective source violates the 
classical energy conditions near the core but tends to restore them 
asymptotically.

In summary, the $\kappa$-dependent correction acts as an effective, 
anisotropic fluid that smoothly interpolates between a de~Sitter-like core and 
an asymptotically vacuum exterior.   Near $r=0$ the negative energy density and 
the corresponding pressure $p_r=-\rho$ 
regularize the central region, removing the curvature singularity, while for 
large $r$ the stress-energy decays exponentially as \(\mathrm{sech}(\pi\kappa 
r^2)\). This behavior is consistent with a non-classical, 
quantum-gravity-induced core, where the violation of energy conditions is the 
price paid for achieving a regular black hole interior. In the standard  limit 
$\kappa\to0$, the effective stresses vanish and the standard Schwarzschild 
solution is recovered.

\subsection{Logarithmic corrected entropy}

Logarithmic corrections to the Bekenstein-Hawking entropy naturally arise when 
quantum or statistical fluctuations of the horizon degrees of freedom are taken 
into account. While the leading-order entropy is proportional to the horizon 
area, subleading corrections appear once quantum fields, quantum geometry, or 
thermal fluctuations near the horizon are included. A general and widely 
encountered form of the corrected entropy is
\begin{equation}
S_{\rm log} = S_{BH} + \lambda \, \ln S_{BH} \ ,
\end{equation}
where $\lambda$ depends on the underlying quantum gravity framework. These 
logarithmic corrections are remarkably universal: they have been derived in 
loop 
quantum gravity \cite{Kaul:2000kf}, in string theory \cite{Sen:2013ns}, in the 
quantum 
geometry approach \cite{Carlip:2000nv}, and in treatments based on thermal 
fluctuations in canonical ensembles \cite{Das:2001ic}. Physically, the 
logarithmic term reflects fluctuations around the classical equilibrium 
configuration of the horizon and becomes especially relevant for small black 
holes or near-extremal configurations. Such corrections can modify 
thermodynamic 
stability and phase behavior, providing an important probe into the microscopic 
origin of gravitational entropy.

For this entropy modification, the corrected metric function  \eqref{Generalfr} 
becomes
\begin{equation}
    f_{\rm log}(r) = 1-\frac{2M \, \pi \, r}{\lambda +\pi  r^2} \ ,
\end{equation}
while the Einstein tensor components  (\ref{rhoeq})-(\ref{ptreq}) are
\begin{eqnarray}
    G_t^t &=&G_r^r = -\frac{4 \pi  \lambda  M}{r \left(\lambda +\pi  
r^2\right)^2} \\  G_\theta^\theta &=& G_\phi^\phi = \frac{2 \pi  \lambda  M 
\left(3 \pi  r^2-\lambda \right)}{r \left(\lambda +\pi  r^2\right)^3} \ ,
\end{eqnarray}
and the gravitational force   \eqref{New} reads
\begin{equation}
    \vec{F}_G=  -2\pi M \frac{S_{\rm log}''(r)}{S_{\rm log}'(r)^2}\hat{r}=  
\frac{\pi  M \left(\lambda -\pi  r^2\right)}{\left(\lambda +\pi  
r^2\right)^2}\hat{r} \ .
\end{equation}
The effective stress-energy tensor components (\ref{rhoeq})-(\ref{ptreq}) 
 become
\begin{eqnarray}
    \rho = -p_r = \frac{\lambda M}{2 r \left(\lambda + \pi r^2\right)^2},
\end{eqnarray}
and 
\begin{eqnarray}
     p_t = \frac{\lambda M \left(3 \pi r^2 - \lambda \right)}{4 r \left(\lambda 
+ \pi r^2\right)^3} \ .
\end{eqnarray}
Thus, the stress tensor clearly shows an anisotropic pressure structure with \( 
p_r \neq p_t \), and a negative radial pressure similar to that found in 
de~Sitter-like cores.   NEC and 
WEC are satisfied. Concerning SEC, since we find that 
    \[
        \rho + p_r + 2p_t = \frac{\lambda M \left(3 \pi r^2 - \lambda 
\right)}{2 
r (\lambda + \pi r^2)^3},
    \]
    which can become negative for \(r^2 < \lambda / (3\pi)\), we conclude that  
SEC is
violated near the core.

In summary, the effective matter source corresponds to an anisotropic fluid 
that mimics a 
regularized gravitational core. At small \(r\), the density approaches the 
finite 
$
\rho(r\to0) \sim \frac{M}{2 \pi \lambda r},
$
while at large \(r\)  it decreases as \(r^{-3}\).  
The negative radial pressure and SEC violation near the origin are signatures 
of a de~Sitter-like vacuum behavior, ensuring a regular interior geometry. 
Thus, this stress-energy distribution effectively describes a smooth transition 
from a quantum-gravity-inspired core to an asymptotically Schwarzschild 
exterior.

\subsection{Entropy in the context of Loop Quantum Gravity (LQG)}

In the context of Loop Quantum Gravity (LQG), non-extensive statistical 
mechanics gives the following modified entropy law 
\cite{Czinner:2015eyk,Nojiri:2022ljp}:
 \begin{equation}
S_{LQG}(A) = \frac{1}{(1-q)} \exp \left[ \frac{(1-q)\Lambda(\gamma_0)A}{4} - 1 
\right],
\end{equation}
where $\Lambda(\gamma_0) = \ln 2 / (\sqrt{3}\pi \gamma_0)$, with $\gamma_0$ the 
Barbero-Immirzi parameter which measures the size of area quanta in Planck 
units, determined by counting
the number of spin-network states corresponding to an event horizon of area 
$A$. 
The corrected metric is 
\begin{eqnarray}
    f_{\rm LQG} (r) =  1-\frac{2 M e^{\pi  \Lambda  (q-1) r^2+1}}{\Lambda  r},
\end{eqnarray}
the components of the Einstein tensor are 
\begin{equation}
    G_t^t = G_r^r = -\frac{4 \pi  M (q-1) e^{\pi  \Lambda  (q-1) r^2+1}}{r} \ \ 
\ \ \ \ \ \ 
\end{equation}
 and
{\small{\begin{equation}
    G_\theta^\theta= G_\phi^\phi = -\frac{2 \pi  M (q\!-\!1) e^{\pi  \Lambda  
(q\!-\!1) 
r^2+1} \left[2 \pi  \Lambda  (q\!-\!1) r^2+1\right]}{r},
\end{equation}}}
while the gravitational force for this case is written as
\begin{equation}
    \vec{F}_G=  -2\pi M \frac{S_{\rm LQG}''(r)}{S_{\rm LQG}'(r)^2}\hat{r}= 
\frac{M  \left[2 \pi  \Lambda  (q-1) r^2-1\right]}{\Lambda \,e^{-\left(\pi  
\Lambda  (q-1) r^2+1\right)} r^2} \hat{r} \ . 
\end{equation}
From the components of the effective stress-energy tensor we can easily compute 
the corresponding fluid parameters as
\begin{eqnarray}
    \rho &=& -p_r = \frac{M (q-1) e^{\pi  \Lambda  (q-1) r^2+1}}{2 r} \ , \\
    p_t &=&  -\frac{M (q-1)  \left[2 \pi  \Lambda  (q-1) r^2+1\right]}{4 
e^{-\left[\pi  \Lambda  (q-1) r^2+1\right]} r},
\end{eqnarray}
thus the matter source behaves as an anisotropic fluid. The radial WEC  is 
satisfied, while for the tangential component we have
    \[
        \rho + p_t = \frac{M (q-1)  \pi  \Lambda  (q-1) r^2}{2 e^{-\left(\pi  
\Lambda  (q-1) r^2+1\right)} r} > 0,
    \]
    for \(\Lambda (q-1) > 0\). Hence, the WEC is satisfied in the physical 
region. Additionally,  NEC coincides with WEC and it is also satisfied. 
Finally, concerning  SEC  we find
    \[
        \rho + p_r + 2p_t = -\frac{M (q-1)  \left[2 \pi  \Lambda  (q-1) 
r^2+1\right]}{2e^{-\left[\pi  \Lambda  (q-1) r^2+1\right]} r},
    \]
    which becomes negative near the core (\(r\to0\)), indicating SEC violation.

In summary, the stress-energy distribution corresponds to an anisotropic matter 
source whose density increases exponentially with \(r^2\) for positive 
\(\Lambda(q-1)\). The negative radial pressure and the SEC violation near 
\(r=0\) imply the presence of a repulsive core that regularizes the central 
region, similar to a de-Sitter vacuum. For large \(r\), the exponential factor 
dominates and the energy density decays rapidly, leading to an asymptotically 
vacuum configuration. Thus, this effective source describes a smooth transition 
between a regular quantum-inspired core and an exterior Schwarzschild-like 
regime.

\subsection{Exponentially corrected entropy}

Exponential corrections to the Bekenstein-Hawking entropy offer an alternative 
approach to encoding possible quantum or statistical modifications to black 
hole thermodynamics. Unlike logarithmic corrections, which typically arise from 
quantum fluctuations, exponential corrections are motivated by non-perturbative 
or holographic effects that become significant near the Planck scale 
\cite{Medved:2004tp, 
Nojiri:2005sr, Pourhassan:2017qrq, Chatterjee:2020iuf}. The 
corrected entropy is   written as
\begin{equation}
S_{\rm exp} = S_{BH} + \eta \, e^{- S_{BH}} \ ,
\label{exp_corr}
\end{equation}
where $\eta$ is a model-dependent constant characterizing the strength and 
scale 
of the correction. Since the correction term is exponentially suppressed for 
large horizon area, classical black holes with large entropy are essentially 
unaffected, ensuring consistency with general relativity in the semiclassical 
regime. However, for small black holes (or near-extremal configurations), the 
exponential contribution can become non-negligible and affect the thermodynamic 
quantities such as heat capacity and free energy. These corrections have been 
studied in contexts including quantum tunneling methods, modified gravity 
theories, and non-perturbative quantum gravity models \cite{Medved:2004tp, 
Nojiri:2005sr, Pourhassan:2017qrq, Chatterjee:2020iuf}. Their impact is 
particularly important in examining the late stages of black hole evaporation 
and the possible resolution of the final state problem.

In this case, the corrected metric is 
\begin{eqnarray}
    f_{\rm exp} (r) = 1-\frac{2 M}{r-\eta  e^{-\pi  r^2} r},
\end{eqnarray}
the Einstein tensor components are
\begin{equation}
    G_t^t = G_r^r = \frac{4 \pi  \eta  M e^{\pi  r^2}}{r \left(e^{\pi  
r^2}-\eta 
\right)^2},
\end{equation}
and 
\begin{equation}
   G_\theta^\theta = G_\phi^\phi = \frac{-2 \pi  \eta  M  \left[\eta +2 \pi  
\eta  r^2+e^{\pi  r^2} \left(2 \pi  r^2-1\right)\right]}{e^{-\pi  r^2}\,r 
\left(e^{\pi  r^2}-\eta \right)^3} \ ,
\end{equation}
and the gravitational force   reads as
\begin{equation}
    \vec{F}_G=  -2\pi M \frac{S_{\rm exp}''(r)}{S_{\rm exp}'(r)^2}\hat{r}= 
-\frac{M e^{\pi  r^2} \left[\kappa  \left(2 \pi  r^2\!-\!1\right)+e^{\pi  
r^2}\right]}{r^2 \left(e^{\pi  r^2}\!-\!\kappa \right)^2} \hat{r} \ . 
\end{equation}
Furthermore, the effective stress-energy tensor components become
\begin{eqnarray}
    \rho &=& -p_r = -\frac{\eta M e^{\pi r^2}}{2 r \left(e^{\pi r^2}-\eta 
\right)^2}, \\
    p_t  &=& -\frac{\eta M  e^{\pi r^2}\left[\eta +2 \pi  \eta  r^2+e^{\pi  
r^2} 
\left(2 \pi  r^2-1\right)\right]}{4 r \left(e^{\pi  r^2}-\eta \right)^3}.
\end{eqnarray}
Therefore, the stress-energy tensor describes an anisotropic fluid with a 
de~Sitter-like 
core, where the radial and tangential pressures differ (\(p_r \neq p_t\)). The 
WEC and NEC are satisfied for large \(r\), while for the SEC we find
\[
        \rho + p_r + 2p_t = -\frac{\eta M e^{\pi r^2}\left[\eta + 2 \pi \eta 
r^2 
- e^{\pi r^2}(1 - 2\pi r^2)\right]}{2 r \left(e^{\pi r^2}-\eta \right)^3},
    \]
    which can become negative near \(r=0\), showing SEC violation in the core 
region. 

In summary, the effective matter source represents a regular anisotropic fluid 
distribution. For small \(r\)  the density remains finite as
\[
\rho(r \to 0) \approx \frac{\eta M}{2 r (\eta - 1)^2} \ ,
\]
and for large \(r\)  it falls off exponentially due to the \(e^{\pi r^2}\) term 
in the denominator. The negative radial pressure and the SEC violation near the 
origin indicate a de-Sitter-like vacuum behavior, which prevents curvature 
singularities. This configuration thus provides a regular black-hole model 
interpolating between a finite-density quantum core and an asymptotically 
Schwarzschild regime.

\section{Conclusions}
 \label{Conclusions}

There is well-known connection  between gravity and thermodynamics, which can 
offer  a useful perspective on the microscopic origin of 
spacetime geometry. Since the identification of black holes as thermodynamic 
systems possessing temperature and entropy, it has become clear that 
gravitational dynamics can be viewed as emergent phenomena arising from 
underlying statistical degrees of freedom. Within this framework, the 
gravitational field equations may be interpreted as thermodynamic relations 
between quantities defined on the horizon, and thus modifications in entropy 
can 
be expected to induce corresponding modifications in the geometry itself.

In the literature, various generalizations of the Bekenstein-Hawking entropy 
have been proposed, motivated by quantum gravitational, statistical, and 
non-extensive frameworks. In particular, Barrow, Tsallis-Cirto, Rényi, 
Kaniadakis, logarithmic, power-law, exponential  and other entropy formulations 
introduce deviations from the standard area law, describing  possible quantum 
or non-equilibrium effects near the horizon. Each of these modified entropies 
carries distinct implications for the thermodynamic stability, phase structure, 
and geometric regularization of black holes.

In this work, we have constructed a general framework linking a modified 
entropy 
function directly to a modified spacetime metric and, subsequently, to an 
effective matter sector of entropic origin. Starting from the first law of 
black-hole thermodynamics, we established an explicit correspondence between 
the 
entropy derivative and the metric function, which naturally leads to an 
emergent 
stress-energy tensor representing an anisotropic effective fluid. This 
procedure 
was then applied to a wide class of entropy models, allowing us to derive the 
associated effective matter sectors, and analyze their physical properties and 
energy conditions in a unified manner.

Although our analysis is performed within the framework of general relativity, 
assuming the Einstein field equations, deviations from Bekenstein-Hawking 
entropy can be interpreted in two complementary ways. They may be viewed either 
as effective matter sources when incorporated into the energy-momentum side of 
the Einstein field equations or, equivalently, as signatures of modified 
gravity 
when incorporated into the geometric side of the Einstein field equations. 
 That being said, the definitive answer or the final interpretation remains 
open, and the correspondence between entropy and geometry allows for a dual 
viewpoint. However, the precise nature of this interpretation is not central to 
the present analysis. The essential result is that entropy itself governs the 
modification of spacetime geometry. Remarkably, the gravitational field 
equations need not be assumed a priori; rather, the geometry can be inferred 
directly from entropic considerations. This provides further evidence in favor 
of gravity as an emergent phenomenon.

Future work can extend this correspondence toward dynamical and cosmological 
spacetimes, studying how generalized entropy functions modify the Friedmann 
equations and the cosmological evolution. Furthermore, the framework can be 
used to investigate the thermodynamic origin of dark energy, to test 
consistency with quantum-gravity considerations, and to confront the results 
with astrophysical observations. Moreover, one could examine the connection of 
modified entropies to modified geometry and gravity. These interesting projects 
are currently under 
investigation.

\section*{Acknowledgements}

Ankit Anand is financially supported by the Institute's postdoctoral fellowship 
at 
IITK. S.D. acknowledges the financial support provided by DST vide Grant No. 
DST/INSPIRE Fellowship/2020/IF200537. A.A. and S.D. acknowledge fruitful 
discussions with Ayan Chatterjee and Anshul Mishra. K.J. thanks Sunny Vagnozzi 
for insightful and fruitful discussions. E.N.S. gratefully acknowledges  the 
contribution of 
the LISA Cosmology Working Group (CosWG), as well as support from the COST 
Actions CA21136 -  Addressing observational tensions in cosmology with 
systematics and fundamental physics (CosmoVerse)  - CA23130, Bridging 
high and low energies in search of quantum gravity (BridgeQG)  and CA21106 -  
 COSMIC WISPers in the Dark Universe: Theory, astrophysics and 
experiments (CosmicWISPers).

\appendix
\section{Horizon Thermodynamics and constraints on $f(r)$}\label{Appn:Horizon 
Thermodynamics and Constraint on $f(r)$}

In this Appendix we show that  starting from the 
Bekenstein-Hawking entropy, it is straightforward to verify that the 
corresponding metric function reproduces the Schwarzschild solution. 
 We start with  metric \eqref{Metric_ansatz}, namely  
\begin{equation}\label{Metric_ansatzA}
    f(r) = 1 - M\,g(r) \, ,
\end{equation}
whose   event horizon $r=r_{+}$ is 
defined by $f(r_{+}) = 0$. We consider  the Bekenstein-Hawking entropy
$
    S  = \pi r_{+}^{2}$ and the Hawking temperature, determined by the 
surface gravity $\kappa = 
f'(r_{+})/2$, namely 
$    T =   \frac{f'(r_{+})}{4\pi} $.
Using the first law of black hole thermodynamics $dM = T\,dS$, we have
\begin{equation}
    \frac{dM}{dr_{+}} = T\,\frac{dS}{dr_{+}} =   \frac{r_{+}}{2}\, 
f'(r_{+}) \ .
\end{equation}
Assuming that the metric function $f(r)$ depends on the mass parameter $M$, the 
horizon condition $f(r_{+},M)=0$ determines $M$ as a function of $r_{+}$, 
i.e.  $M = M(r_{+})$. Differentiating the horizon condition implicitly gives
\begin{equation}
    \frac{dM}{dr_{+}} = -\frac{\partial_{r} f(r_{+},M)}{\partial_{M} 
f(r_{+},M)},
\end{equation}
and equating this with the thermodynamic relation above yields the general 
consistency condition
\begin{equation}
    -\frac{\partial_{r_+} f(r_{+},M)}{\partial_{M} f(r_{+},M)} = 
\frac{r_{+}}{2}\, f'(r_{+}) \ .
\end{equation}
This relation provides a constraint on the admissible metric functions $f(r)$ 
for a black hole satisfying the first law with the Bekenstein-Hawking entropy. 
Starting with the simple power-law ansatz $f(r)=1-c\,M\,r^{-p}$, imposing the 
above condition for arbitrary $r_{+}$ uniquely selects $p=1$ and $c=2$, 
recovering the Schwarzschild form
\begin{equation}
    f(r)=1-\frac{2M}{r} \ .
\end{equation}
Thus, within this class of metrics, the  Schwarzschild one is singled out solely 
by the 
requirement that the first law $dM=T\,dS$ holds together with the Bekenstein 
entropy formula.  Finally, we note that the uniqueness of the 
Schwarzschild solution obtained here follows from the imposed symmetry 
assumptions together with the thermodynamic constraint.

\bibliographystyle{utphys}
\bibliography{ref}

\end{document}